\documentstyle[12pt]{article}
\if@twoside  m
\else
\fi
\newcommand{\ie}{{\em i.e.}}
\newcommand{\eg}{{\em e.g.}}
\newcommand{\cf}{{\em cf. }}

\newcommand{\lhs}{{\em lhs }}

\newcommand{\R}{I\!\!R}
\newcommand{\OO}{{\cal O}}

\begin{document}
\title{Electron trapping by a current vortex}
\date{}
\author{F. Bentosela,$^{a,b}$ P. Exner,$^{c,d}$ and 
V.A. Zagrebnov$^{a,b}$}
\maketitle
\begin{quote}
{\small \em a) Centre de Physique Th\'{e}orique, C.N.R.S., 
F--13288 Marseille Luminy \\ 
b) Universit\'{e} de la Mediterran\'{e}e (Aix--Marseille II),
F--13288 Luminy  
\\ c) Nuclear Physics Institute, 
Academy of Sciences, CZ--25068 \v Re\v z near Prague, \\
d) Doppler Institute, Czech Technical University, B\v rehov{\'a} 7,
CZ-11519 Prague, \\ 
\rm \phantom{e)x}bentosela@cpt.univ-mrs.fr, exner@ujf.cas.cz,
zagrebnov@cpt.univ-mrs.fr}

\end{quote}

\begin{abstract}
\noindent
We investigate an electron in the plane interacting with the magnetic
field due to an electric current forming a localized rotationally
symmetric vortex. We show that independently of the vortex profile an
electron with spin antiparallel to the magnetic field can be trapped if the
vortex current is strong enough. In addition, the electron scattering on
the vortex exhibits resonances for any spin orientation. On the other
hand, in distinction to models with a localized flux tube the present
situation exhibits no bound states for weak vortices.
\end{abstract}


\noindent
Interaction of charged particles with a localized magnetic field has been 
a subject of interest for a long time, both from the theoretical and
experimental point of view --- see, \eg, \cite{GBG,St} and references
therein. Such a field can have different sources, for instance, it may be
induced by an electric current having one or more vortices. A lot of
attention was payed in the last decade to vortex bound states in
superconductors whose dynamics is governed by the Bogoliubov--de Gennes
equation --- \cf \cite{HRD,SHDS,GS} and a bibliography given in 
\cite{HIM}. 

Another, much simpler example involves a Pauli electron interacting
with a flux tube modelling a vortex magnetic field --- it is is appealing,
in particular, since it has been observed that vortices appear often in the
probability current associated with mesoscopic transport --- see \cite{ESF}
and the literature there. The system of a tube and an electron has been
investigated in a fresh paper by Cavalcanti et al. \cite{CFC} who, however,
seem to be unaware of another recent studies of the problem --- \cf
\cite{Mo} and references therein. In these papers, the field is assumed to
be constant within a circle and zero otherwise --- the conclusion is
then that in one spin state the electron can be always trapped by the
vortex, independently of the magnetic flux value, as long as the effective
gyromagnetic factor $\,g^*>2\,$. This covers the physically important
case of a free electron with $\,g^*=2.0023$.

However, this claim depends substantially on the used magnetic field
Ansatz. To illustrate this point we analyze in this letter the situation
where the {\em vortex current distribution} represents the input. On one
hand we are able to generalize the result of \cite{Mo,CFC} by showing that
vortices can trap electrons independently of their profile, and what is
physically equally important, that they cause resonances in the
electron--vortex scattering which get sharper as the vortex strength
grows. On the other hand, the trapping needs in this more realistic
setting certain minimum vortex strength (measured  by the total
circulating current or, say, the dipole moment of the field) to occur. To
make things simpler we suppose that the vortex is {\em centrally
symmetric.} We are convinced that the symmetry is not vital for the
conclusions, but its absence makes the treatment technically much more
complicated and we postpone the discussion of the general case to another
paper.

The dynamics of a nonrelativistic electron in the plane exposed to a
perpendicular but nonhomogeneous magnetic field is given by the
respective Pauli equation, \ie, by the Hamiltonian
\begin{equation} \label{Pauli full}
H\,=\,{1\over 2m^*}\, \left( -i\hbar\vec\nabla\,+\,{e\over c}
\vec A(\vec x)\right)^2+\, {1\over 2}\, g\mu_B \sigma_3 B(\vec x)\,,
\end{equation}
where $\,m^*$ is the effective electron mass, $\,g\,$ the gyromagnetic
factor, $\,\mu_B= e\hbar/2m_ec\,$ the Bohr magneton, and the sign choice 
corresponds to the negative charge $\,-e\,$ of the electron. In the
rational units, $\,2m^*= \hbar= c= 1\,$ this becomes 
\begin{equation} \label{Pauli}
H\,=\,\left( -i\vec\nabla\,+\,e\vec A(\vec x)\right)^2+\, 
{1\over 2}\,g^*e\sigma_3 B(\vec x)
\end{equation}
with the effective gyromagnetic factor $\,g^*= g\,{m^*\over m_e}\,$. Since
$\,H\,$ is matrix--diagonal, each spin state can be treated separately.
Moreover, we shall consider the situation when the magnetic field is
rotationally symmetric. This allows us to perform the partial--wave
decomposition and to replace $\,H\,$ by the family of operators
\begin{equation} \label{partial wave Hamiltonian}
H_{\ell}^{(\pm)}=\,-\,{d^2\over dr^2} \,-\,{1\over r}\, {d\over dr}
\,+\, V_{\ell}^{(\pm)}(r)\,, \quad V_{\ell}^{(\pm)}(r):=\, 
\left( eA(r)+\, {\ell\over r} \right)^2 \pm\, {1\over 2}\,g^*e B(r)
\end{equation}
on $\,L^2(\R^+,r\,dr)\,$. To determine their spectral properties, we have
to specify the involved functions, the angular component $\,A(r)\,$ of the
vector potential and the related magnetic field, $\,B(r)= A'(r)+
r^{-1}A(r)\,$.

We have said that they correspond a circulating electric current in the
plane, which is supposed to be anticlockwise and having the angular
component only, $\,\vec J(\vec x)= \lambda \delta(z) J(r) \vec
e_{\varphi}\,$.  Here $\,r,\varphi\,$ are the polar coordinates in the
plane; the total current is $\,\lambda \int_0^{\infty} J(r)\,dr\,$. 
The positive parameter $\,\lambda\,$ can be, of course, absorbed into
$\,J\,$;  we introduce it as a tool to control the vortex ``strength". The
current density is supposed to obey the following modest requirements:
\begin{description}
\vspace{-.6ex}
\item{\em (i)} $\;J\,$ is $\,C^2$ smooth and non--negative, 
$\,J(r)\ge 0\,$,
\vspace{-1.2ex}
\item{\em (ii)} at the origin $\,J(r)= ar^2+\OO(r^3)\,$,
\vspace{-1.2ex}
\item{\em (iii)} at large distances $\,J(r)= \OO(r^{-3-\epsilon})\,$ 
for some $\,\epsilon>0\,$.
\vspace{-.6ex}
\end{description}
The vector potential in the plane $\,z=0\,$ is then also anticlockwise;
its magnitude
\begin{equation} \label{A}
A(r)\,=\, \lambda\, \int_0^{\infty} dr' r' J(r')\, \int_0^{2\pi}\, 
{\cos\varphi'\, d\varphi' \over \left(r^2\!+ r'^2\!- 2rr'\cos\varphi' 
\right)^{1/2}}
\end{equation}
is obtained by summing the contribution from all the circular current 
lines \cite{Ja},
\begin{equation} \label{A2}
A(r)\,=\, \lambda\, \int_0^{\infty} \,J(r')\, {4r\over r+r'}\, 
{(2-\rho^2)K(\rho^2)- 2E(\rho^2) \over \rho^2}\:dr'\,,
\end{equation}
where $\,\rho^2:= {4rr'\over (r+r')^2}\,$ and $\,K,\,E\,$ are the full
elliptic integrals of the first and the second kind, respectively. Using
\cite[17.3.29,30]{AS} (pay attention to a misprint there), one can 
also cast (\ref{A2}) into the form
\begin{equation} \label{A3}
A(r)\,=\, 4\lambda\, \int_0^{\infty} J(r')\, {r'\over r_<}\: 
\left\lbrack K\left(r_<^2\over r_>^2 \right) - 
E\left(r_<^2\over r_>^2\right)\right\rbrack\,dr'\,,
\end{equation}
where we have used the usual shorthands, $\,r_<:= \min(r,r')\,$ and
$\,r_>:= \max(r,r')\,$; the same result follows directly from (\ref{A})
and \cite[3.674.3]{GR}. In view of {\em (i)} the integral is finite for
every $\,r\,$, because $\,E(\zeta)\,$ is regular at $\,\zeta=1\,$ and
$\,K(\zeta)\,$ has a logarithmic singularity there.

Let us denote the Pauli Hamiltonian (\ref{Pauli}) with the vector
potential (\ref{A3}) by $\,H(\lambda)\,$; the symbol
$\,H_{\ell}^{(\pm)}(\lambda)\,$ will be used for its spin and orbital
momentum components. Our main result is then the following:
\vspace{3mm}

\noindent
{\bf Theorem.} Under the stated assumptions, $\,\sigma(H(\lambda))=
[0,\infty)\,$ for $\,|\lambda|\,$ small enough. On the other
hand, $\,H_0^{(-)}(\lambda)\,$ has a negative eigenvalue for a
sufficiently large $\,\lambda\,$.
\vspace{2mm}

\noindent
{\em Proof:} The first claim has to be checked for any
$\,H_{\ell}^{(\pm)}(\lambda)\,$. By {\em (i)}, the potentials
$\,V_{\ell}^{(\pm)}\,$ are $\,C^1$ smooth; they decay at infinity as 
\begin{equation} \label{decay}
V_{\ell}^{(\pm)}(r)\,=\, {\ell^2\over r^2}\,+\, \lambda e m\, 
{2\ell\mp g^*\over r^3}\,+\, \OO(r^{-3-\epsilon})\,,
\end{equation}
where $\,m:= \pi\, \int_0^{\infty} J(r')\, r'^2 dr'\,$ is the dipole moment
of the current for $\,\lambda=1\,$ --- \cf \cite{Ja}. Consequently,
$\,\sigma_{ess}(H_{\ell}^{(\pm)}(\lambda))= [0,\infty)\,$ following
\cite[Sec. XIII.4]{RS}. We rewrite the potentials into the form
\begin{equation} \label{lambda dependence}
V_{\ell}^{(\pm)}(r)\,=\, \left( \lambda eA_1(r)+\, {\ell\over r} 
\right)^2 \pm\, {\lambda\over 2}\,g^*e B_1(r)\,,
\end{equation}
where the the indexed magnetic field refers to the value 
$\,\lambda=1\,$. Since $\,H_{\ell}^{(\pm)}(\lambda)\,$ is just the
$s$--wave part of the two--dimensional Schr\"odinger operator with 
the centrally symmetric potential (\ref{lambda dependence}), it is
sufficient to check that the latter has no negative eigenvalues. 
If $\,\ell=0\,$ the potential decay allows us to apply Thm.~3.4 of 
\cite{Si} by which a negative eigenvalue exists for small positive 
$\,\lambda\,$ if and only if $\,\int_0^{\infty} V_{\ell}^{(\pm)}(r)\,
r\,dr \le 0\,$. However, the flux through the circle of radius $\,r\,$ is
$\,2\pi \int_0^r B(r')\,r'dr'= 2\pi r A(r)\,$, so the second term
in (\ref{lambda dependence}) does not contribute and the integral is
determined by the first one which is positive for any nonzero
$\,\lambda\,$. If $\,\ell\ne 0\,$ the decay is too slow, but this 
difficulty is easily overcome. We replace the first term, \eg, by 
$\,\left(\lambda eA_1(r)+\, {\ell\over r} \right)^2 \Theta(r_0\!-r)\,$
with a positive $\,r_0\,$ and obtain the absence of a negative eigenvalue
for a small $\,\lambda\,$; the same is by the minimax principle true for
the original operator.

For the existence claim the behaviour of the potential around the origin
is vital. We shall write the vector potential in the form
\begin{equation} \label{A small}
A(r)\,=\, \lambda\mu r + \alpha_0(r)\,, \qquad \mu\,:=\, \int_0^{\infty} 
J(r')\, {dr'\over r'}\,.
\end{equation}
The behaviour of $\,\alpha_0\,$ follows from (\ref{A3}) and the following
estimate on the difference of the elliptic integrals,
\begin{equation} \label{elliptic}
{\pi\zeta\over 4}\,\le\, K(\zeta)-E(\zeta) \,\le\, 
{\pi\over 2}\, \left( {\zeta\over 8}\,-\, {3\over 8} \ln (1-\zeta) 
+ b\zeta^2 \right)
\end{equation}
for a sufficiently large $\,b>0\,$, which is a straightforward consequence
of \cite[17.3.11,12]{AS} and \cite[8.123.2]{GR}. On the lower side we
thus get
$$
\alpha_0(r)\,\ge\, {\pi\over r^2}\, \int_0^r J(r')\, r'^2 dr'\,-\, 
\pi r\, \int_0^r J(r')\,{dr'\over r'}\,\ge\, 
-\,{3\over 4}\, \pi ar^2 + \OO(r^3)\,. 
$$
The upper bound is similar. In view of {\em (i)--(iii)}, $\,J(r)\le \tilde
ar\,$ for a suitable $\,\tilde a\,$; in the logarithmic term we employ the
Taylor expansion which gives
$$
{3\pi\tilde a\over 4r}\, \int_r^{\infty}\, \sum_{j=2}^{\infty}\,
{r^{2j}\over  (r')^{2j-2}}\,dr' \,=\, {3\pi\tilde a\over 4}\,r^2 
\sum_{j=2}^{\infty}\, {1\over j(2j\!-\!3)}\,.
$$
Together we find $\,\alpha_0(r)= \OO(r^2)\,$. This further implies 
\begin{equation} \label{B small}
B(r)\,=\, 2\lambda\mu + \beta_0(r)\,, \qquad \beta_0(r)\,:=\, 
\alpha_0'(r)+\, {1\over r} \alpha_0(r)\,\in\OO(r)\,.
\end{equation}
The operator $\,H_0^{(-)}(\lambda)\,$ can be therefore written as
\begin{equation} \label{s-wave Hamiltonian}
H_0^{(-)}(\lambda)\,=\,-\,{d^2\over dr^2} \,-\,{1\over r}\, 
{d\over dr}\,+\, \left(\lambda e(\mu r+\alpha_0(r))\right)^2
-\, {1\over 2}\, \lambda e g^*(2\mu +\beta_0(r))\,.
\end{equation}
Using the rescaled variable $\,u:= r\sqrt{\lambda}\,$ we find it to be
unitarily equivalent to the operator
\begin{equation} \label{equivalent operator}
\lambda A_{\lambda} \qquad {\rm with} \qquad A_{\lambda}= A_0+ W_{\lambda}
\end{equation}
on $\,L^2(\R^+,u\,du)\,$, where
\begin{equation} \label{A_0}
A_0\,:=\,-\,{d^2\over du^2} \,-\,{1\over u}\, {d\over du}
\,- g^*e\mu + e^2\mu^2 u^2
\end{equation}
and
\begin{equation} \label{W}
W_{\lambda}(u)\,:=\,2\sqrt{\lambda} e^2\mu\, u\alpha_0\left(u\over
\sqrt{\lambda} \right)\,+\, \lambda e^2\, \alpha_0^2\left(u\over
\sqrt{\lambda} \right)\,-\, {1\over 2}\, g^*e\, \beta_0\left(u\over
\sqrt{\lambda} \right)\,.
\end{equation}
We have clearly $\,\sigma_{ess}(A_{\lambda})=
\sigma_{ess}(\lambda A_{\lambda})= [0,\infty)\,$ for any $\,\lambda>0\,$,
while $\,A_0\,$ as the $s$--wave part of the two--dimensional harmonic
oscillator has a purely discrete spectrum. Despite the fact that the
perturbation is large (the maximum of $\,W_{\lambda}\,$ grows linearly
with $\,\lambda\,$), one may attempt to employ the {\em asymptotic}
perturbation theory. 

Unfortunately, Thms.~VIII.3.11,13 of \cite{Ka} cannot be applied directly,
because the family $\,\{W_{\lambda}\}\,$ is not monotonous. Instead we
use the fact that $\,W_{\lambda}\,$ tends to zero pointwise as
$\,\lambda\to\infty\,$, since the above estimates yield
\begin{equation} \label{W estimate}
\left|W_{\lambda}(u)\right|\,\le\, \left(2 e^2\mu\, c_{\alpha} u^3 
+\, {1\over 2}\, g^*e\, c_{\beta} u \right) \lambda^{-1/2} \,+\, 
e^2 c_{\alpha}^2 u^4 \lambda^{-1}
\end{equation}
for some positive $\,c_{\alpha},\, c_{\beta}\,$. The family
$\,\{A_{\lambda}\}\,$ can be estimated from below because the potentials
have in view of (\ref{W}) a uniform lower bound. Hence one can choose
$\,\xi_0<0\,$ which belongs to $\,\rho(A_{\lambda})\,$ for all
$\,\lambda\,$, and the resolvents form a bounded family of positive
operators, $\,(A_{\lambda}-\xi)^{-1}\le (\xi_0\!-\xi)^{-1}\,$, for any
$\,\xi<\xi_0\,$. Next we use a trick based on the resolvent identity:
for a vector $\,f=(A_0-\xi)g\,$ with a fixed $\,g\in C_0^{\infty}(\R^+)\,$
we have
\begin{equation} \label{resolvent estimate}
\left\|(A_{\lambda}-\xi)^{-1}f -(A_0-\xi)^{-1}f \right\| \,=\, 
\left\|(A_{\lambda}-\xi)^{-1}W_{\lambda}g \right\| \,\le\, 
(\xi_0\!-\xi)^{-1} \left\|W_{\lambda}g \right\| \,\to\, 0
\end{equation}
as $\,\lambda\to\infty\,$ in view of (\ref{W estimate}) and the compact
support of $\,g\,$. However, the family of such $\,f\,$ is dense in
$\,L^2(\R^+,u\,du)\,$, so $\,A_{\lambda}\to A_0\,$ in the strong resolvent
sense. 

This allows us to employ Thm.~VIII.1.14 of \cite{Ka} by which to any
$\,\nu_n\in \sigma_p(A_0)\,$ there is a family of $\,\nu_n(\lambda)\in
\sigma(A_{\lambda})\,$ such that $\,\nu_n(\lambda)\to\nu_n\,$ as
$\,\lambda\to\infty\,$. It is therefore decisive that the unperturbed
eigenvalue is stable in the sense of \cite{Ka}, which means negative in
our case. The spectrum of $\,A_0\,$ is given explicitly by 
\begin{equation} \label{HO spectrum}
\nu_n\,=\, e\mu\left(4n+2-g^* \right)\,, \qquad n\,=\, 0,1,\dots\,,
\end{equation}
so the condition is satisfied for $\,n=0\,$ if $\,g^*>2\,$ (as in
\cite{CFC}, the next eigenvalue comes to play for $\,g^*>6\,$ which is
physically not appealing). $\quad\Box$ 
\vspace{3mm}

We finish the letter by remarks on extensions of the result and
related topics:
\begin{description}
\item{\em (a)} The difference in the {\em weak--coupling behaviour}
comparing to \cite{Mo,CFC} is not surprising. In their case
$\,\int_0^{\infty} V_0^{(-)}(r)\, r\,dr\,$ is dominated for small
$\,\lambda>0\,$ by the negative term due to the well. In reality, however,
the magnetic field flux lines are closed in $\,\R^3$, so the well is
compensated by a repulsive tail, small but extended, which prevents the
trapping.
\vspace{-1.2ex}
\item{\em (b)} The asymptotic perturbation theory yields also the ground
state behaviour as $\,\lambda\to\infty\,$. By \cite[Thm.~VIII.2.6]{Ka} the
leading--order correction to $\,\nu_0\,$ is
$$
(\psi_0,W_{\lambda}\psi_0)= {3e\over 8} (g^*\!+2) \sqrt{\pi\over e\mu}
\, \alpha_0''(0)\, \lambda^{-1/2} + \OO(\lambda^{-1})\,,
$$
where $\,\psi_0\,$ is the ground--state eigenfuction of the
two--dimensional harmonic oscillator. However, $\,\alpha_0''(0)=0\,$, so
the ground state of the original operator $\,H_0^{(-)}(\lambda)\,$ behaves
as $\,-\lambda e\mu (g^*\!-2) +\OO(1)\,$.
\vspace{-1.2ex}
\item{\em (c)} Large $\,\lambda\,$ give rise to an {\em orbital series} of
bound states: the above also argument works for
$\,H_{\ell}^{(-)}(\lambda)\,$ with $\,\ell=-1,-2,\dots\,$. The potential
in (\ref{A_0}) is replaced at that by $\,e^2\mu^2 u^2 +\ell^2 r^{-2}
+e\mu(2\ell\!-\!g^*)\,$, and one looks for negative eigenvalues among
$\,\nu_{n,\ell}= e\mu \left(4n+ 2(|\ell|\!+\!\ell)+2-g^*\right)\,$. The
critical $\,\lambda\,$ at which the eigenvalue emerges from the continuum
is naturally $\,\ell$--dependent.
\vspace{-1.2ex}
\item{\em (d)} Positive eigenvalues of $\,A_0\,$ are unstable in the sense
that they disappear in the continuum once the perturbation (\ref{W}) is
turned on. Following \cite[Sec.~VIII.5]{Ka}, however, they give rise to
spectral concentration as $\,\lambda\to\infty\,$ which is manifested by
{\em resonances} in electron scattering on the vortex. Knowing the shape
of the potential barrier, one can compute their widths which vanish
exponentially fast with $\,\lambda\,$. For a fixed $\,\ell\,$, the number
of resonances grows asymptotically linearly with $\,\lambda\,$, because
the eigenvalues of the operator (\ref{A_0}) are equally spaced and the top
of the potential barrier in $\,A_{\lambda}\,$ is asymptotically linear in
$\,\lambda\,$. Due to the same reason, resonances exist at large
$\,\lambda\,$ for {\em both spin signs.} Notice also that the existence of
resonances is not restricted by the value of the gyromagnetic factor, and
therefore they may be observed in semiconductor systems where $\,|g^*|$ is
typically less than one, in some cases even of order of $\,10^{-2}$.
\vspace{-1.2ex}
\item{\em (e)} The above mentioned resonance scattering is a {\em purely 
quantum effect.} A classical electron can be, of course, trapped in
the current vortex if it is placed inside the potential barrier
(\ref{partial wave Hamiltonian}) with the energy less than its top
(since there is no spin in this case, $\,g^*=0\,$, the well bottom is
at zero). On the other hand, an electron of the same energy outside
the barrier gets scattered without the possibility of entering
temporarily the interior of the vortex.
\vspace{-1.2ex}
\item{\em (f)} In addition to smooth current distributions discussed
above, the case $\,J(r)= \delta(r\!-\!R)\,$ is of a practical interest.
As an illustration, imagine two adjacent thin films, one supporting a free
electron gas while the other is equipped with a mesoscopic ring in which a
persistent current circulates. The effective potential is now given
explicitly by (\ref{Pauli}) and (\ref{A3}); in distinction to the smooth
case it has a singularity of the type $\,(r\!-\!R)^{-1}\,$ and the
operators $\,H_{\ell}^{(\pm)}(\lambda)\,$ need a regularization. Since the
$\,\delta\,$ function above is an idealization of a sharply localized
distribution, it is reasonable to choose for this purpose from the family
of all admissible procedures a scheme based on the principal value of the
singular potential \cite{NZ,Ku}.
\vspace{-1.2ex}
\item{\em (g)} The reader may wonder whether the preferred spin
orientation of the ground state does not conflict with the second
theorem of Aharonov and Casher \cite{AC}. The present model offers a
good illustration that this claim should be taken {\em cum grano salis}
indeed (in distinction to the first one which is sound --- \cf\cite{Th}).
Recall that the Pauli Hamiltonian (\ref{Pauli}) can be factorized into the
product $\,(\Pi_1\!-i\sigma_3\Pi_2)(\Pi_1\!+i\sigma_3\Pi_2)\,$, where
$\,\vec \Pi:= -i\vec\nabla+\!e\vec A\,$; if $\,\Psi\,$ solves
$\,H\Psi=E\Psi\,$ then $\,\tilde\Psi:= (\Pi_1\!+i\sigma_3\Pi_2)\Psi\,$
solves the equation with spin term sign switched. The nonzero component of
the rotationally symmetric spinor $\,\Psi\,$ turns at that into
$\,-ie^{i\varphi}(\psi'(r)\!-\!eA(r)\psi(r))\,$. For the ground--state
eigenfunction we have $\,\psi(0)\ne 0\,$ while $\,\psi'(0)=0\,$ due to the
smoothess of $\,\Psi\,$. However, the vector potential satisfies (\ref{A
small}) with a positive coefficient. To belong to the domain of $\,H\,$,
the nonzero component of $\,\tilde\Psi\,$ must be smooth at the origin.
This requires $\,\psi''(0)\psi(0)^{-1}= \lambda e\mu\,$ but the \lhs
is negative; recall that $\,\psi\,$ approaches the harmonic--oscillator
ground state for large $\,\lambda\,$.
\vspace{-1.2ex} 
\item{\em (h)} Notice finally another difference. The model with a flux
tube can have a natural ``squeezing limit" in terms of Aharonov--Bohm
Hamiltonians with a pointlike magnetic flux \cite{Mo,AT,DS} because the
vector potentials coincide outside the flux tube and the attractive part
tends to a $\,\delta$--well (with the exception of a single one, the
resonances mentioned in (d) are lost at that as it is usual in such
situations --- \cf\cite{AGHH}). On the other hand, the present case is more
complicated being essentially three--dimensional as far as the magnetic
field is concerned.
\vspace{-1.2ex}
\end{description}
%


\subsection*{Acknowledgment}

P.E. thanks for the hospitality extended to him at C.P.T. where this
work was done. The research has been partially supported by GAAS under the
contract A1048801.

\end{document}